\documentclass[12pt,preprint]{aastex}

\shortauthors{Spangler et al.}
\shorttitle{Dusty Debris Around Solar-Type Stars}

\begin{document}

\title{ Dusty Debris Around Solar-Type Stars: Temporal Disk Evolution }

\author{ C. Spangler\altaffilmark{1} and A. I. Sargent }
\affil{ Division of Physics, Mathematics and Astronomy, California Institute of Technology, MS~105-24, Pasadena, CA 91125 }
\email{ celeste@spanglers.com, afs@astro.caltech.edu }

\and

\author{ M.D. Silverstone\altaffilmark{2}, E.E. Becklin, and B. Zuckerman }
\affil{ Division of Astronomy and Astrophysics, University of California, Los~Angeles, Los~Angeles, CA 90095-1562 }
\email{ murray@as.arizona.edu, becklin@astro.ucla.edu, ben@astro.ucla.edu}

\altaffiltext{1}{present address: 6262 Bothell Cir., San Jose, CA  95123}
\altaffiltext{2}{present address: Steward Observatory, University of Arizona, Tucson, AZ  85721 }

\begin{abstract}

  Using ISO-ISOPHOT we carried out a survey of almost 150 stars to search for
  evidence of emission from dust orbiting young main sequence stars, both in
  clusters and isolated systems.  Over half of the detections are new examples
  of dusty stellar systems, and demonstrate that such dust can be detected
  around numerous stars older than a few $\times 10^6$~years.  Fluxes at
  60\,$\micron$ and either 90 or 100\,$\micron$ for the new excess sources
  together with improved fluxes for a number of IRAS-identified sources are
  presented.  Analysis of the excess luminosity relative to the stellar
  photosphere shows a systematic decline of this excess with stellar age
  consistent with a power law index of $-2$.

\end{abstract}


% if desiring to write a note to the editor
%\notetoeditor{TEXT}

\section{Introduction}

  Circumstellar disks are generally accepted as a natural by-product of star
  formation \citep[e.g.][]{shu93}, and there is considerable observational
  evidence for the presence of such disks around very young solar mass stars
  \citetext{e.g. \citealp{sar96}, and references therein;
  \citealp*{nat00,and00,mun00,wil00}}.  The IRAS detection of far infrared
  radiation from main sequence stars at a level much greater than could be
  attributed to the stellar photospheres was, nonetheless, unexpected
  \citep[e.g.][]{aum84}.  These stars became known as ``Vega-type'' stars,
  after one of the first examples.  This infrared excess radiation was
  interpreted as thermal emission from dust in orbit about the stars
  \citetext{c.f. \citealp{bac93}, and references therein; \citealp*{lag00}}.
  Early coronographic observations of scattered light from the Vega-type star
  $\beta$~Pictoris indicated an edge-on disk morphology \citep{smi84}.  Recent
  imaging of thermal emission at mid-IR and mm wavelengths from Vega-type
  stars, including HR~4796A, MWC~480, and $\epsilon$~Eridani, continue to
  provide support for the disk interpretation
  \citep*{koe98,jay98,hol98,man97,gre98}.  Dramatic disk-like images in
  reflection around HR~4976A and the Herbig Ae/Be star HD~141569 have been
  recently published \citep{sch99,wei99,aug99}.

  Typically, the ages of T-Tauri stars with associated disks are of order
  $10^{6}$~years old or less.  The Vega-type stars have ages ranging from
  $10^{7}$ to $10^{9}$~years.  While known examples of Vega-type systems are
  fairly well studied, their limited number provides little information on the
  evolution of circumstellar dust during these later epochs.  Yet it is these
  later epochs that are of critical importance to understanding the
  development of planetary systems \citep{lag00}.  Gathering a statistical
  sample of intermediate age objects is difficult.  Not only is the time to
  evolve from proto-planetary to planetary debris disks likely to be brief,
  but disk emission drops dramatically with age as grains grow and/or
  dissipate, reducing the emitting surface area.  As planetary embryos form
  close to the stars, the hot inner dust is probably cleared.  There are also
  many mechanisms capable of reorganizing or removing the outer primordial
  circumstellar dust, including grain growth,  Poynting-Robertson drag, and
  radiation pressure, in times short compared to the ages of older Vega-type
  stars.  In fact, for older stars to exhibit dusty disks there must be an
  ongoing source of particles; a favored explanation is that some
  planetesimals are colliding and fragmenting.  There is substantial evidence
  that extrasolar planets are common \citep*{mar00}, implying smaller
  planetesimals, collisions of which could provide a reservoir of
  circumstellar dust, are also common.

  Since thermal emission from the little remaining outer, cooler material
  peaks at wavelengths of $50 - 150\,\micron$, airborne or space-based
  instruments are required to detect intermediate age and older disks.
  However, IRAS, with sensitivities of a few hundred mJy at 60 and
  100\,$\micron$, detected only the nearest and brightest disks -- mostly
  those associated with main sequence stars of spectral type A.  In addition,
  the low spatial resolution of IRAS limited the ability to determine if
  emission was associated with a specific object, particularly at
  100\,$\micron$ where galactic cirrus confusion is significant.  ISO, with a
  factor of two improvement in resolution, $46\arcsec$ pixels at
  $60 - 100\,\micron$, and an order of magnitude increase in sensitivity over
  IRAS, has expanded our capability to search for circumstellar dust.

  Here we describe our ISO  program\footnotemark \ of observations to search
  for dusty circumstellar material, especially around stars of spectral types
  F and G, and also to study the evolution of disks from the pre-main sequence
  through the main sequence stages.  Details of the target selection are
  presented in Section~\ref{sec:source}; an account of observations, data
  reduction and basic results is given in Section~\ref{sec:obsres}; excess
  radiation calculations are described in Section~\ref{sec:excess}.  Results
  and implications of our observations are discussed in
  Section~\ref{sec:disc}, and summarized in Section~\ref{sec:sum}.

  \footnotetext{A NASA/ISO Key Project. }

\section{Target Selection \label{sec:source}}

  Our targets fell into three distinct categories:

  \begin{quote}
  1. Main sequence members of relatively nearby, generally $< 120$~pc,
  open clusters including $\alpha$~Persei, Coma~Berenices, Hyades, Pleiades,
  and the Ursa~Major nucleus and stream. Cluster ages are between 50 and
  700~Myrs and the target stars span spectral types A through K.

  2. Selected classical and weak-line T~Tauri stars in the Chamaeleon~I,
  Scorpius, and Taurus star forming clouds at $\sim 150$~pc.  Some of these
  were detected by IRAS but beyond 60 microns only upper limits to the fluxes
  are available.  

  3. A small sample of relatively nearby, $< 60$~pc, isolated stars with
  indications of youth.
  \end{quote}

  A major problem in defining an evolutionary sequence for circumstellar disk
  characteristics arises because it difficult to assign an age to an isolated
  star.  Age determinations rely on a number of factors including the star's
  position in the HR-diagram, its metallicity, and the depth of its convection
  layer.  Even with well-defined observable parameters, the calculated age can
  vary greatly depending on stellar models used.  For stars in clusters, a
  more accurate age determination is possible based on the main sequence
  turn-off point.  By selecting target objects from well-studied clusters,
  ages are fairly well defined.  In order to encompass a variety of
  evolutionary characteristics in any associated disk, stars in our sample
  clusters range from the pre-main sequence (PMS), ages $\lesssim 10$~Myr, to
  the young main sequence, ages up to 1~Gyr.  Reasons to include pre-main
  sequence stars in our study are to establish the initial characteristics of
  the dusty material whose evolution we wish to understand and to enable an
  assessment of the fraction of weak-line T~Tauri stars (WTTs) which support
  disks.  Weak-line T~Tauri stars are similar to classical T~Tauri stars
  (CTTs), but exhibit only weak H$\alpha$ emission and show no other
  indications of the presence of an accreting disk \citep{cal00}.

  Since the contrast between radiation from the stellar photosphere and the
  disk increases with longer wavelengths, we confined our search to ISO's
  60 and 90/100\,$\micron$ wavelength bands.  IRAS 3$\sigma$ sensitivity
  limits at 60 and 100\,$\micron$ were typically 200 -- 300~mJy and
  500 -- 3000~mJy respectively, depending on the degree of cirrus confusion.
  Pre-launch sensitivity estimates for ISOPHOT, on which we based our original
  target selections, were of order 1~mJy (3$\sigma$ RMS) with a 256~s
  observation in these bands.  Since an F0~V star at 100~pc typically has
  photospheric fluxes of 3.6 and 1.2~mJy at 60 and 100\,$\micron$,
  respectively, we expected photospheric fluxes characteristic of
  near-solar-type stars could be detected out to distances of about 100~pc.
  Beyond this distance and for later-type stars, only fairly strong emission
  in excess of photospheric values would be detected.  Consequently, our
  candidate open clusters needed to be nearby, $\lesssim 100$~pc.  In the late
  stages of star formation, at ages $\lesssim 50$~Myr, disks are more likely
  to be brighter and could be detected up to about 200~pc.  Our choice of
  candidate clusters was also constrained by conflicts with ISO Core programs
  and the ISO pointing restrictions.  Complete surveys were out of the
  question.  Nevertheless, the ISO lifetime was sufficient to enable some
  observations in all clusters we considered interesting.
 
  Table~\ref{tab:clus} lists the clusters from which we selected targets,
  along with their ages, distances, the number of targets observed, observing
  modes, and integration times.

\placetable{tab:clus}

  Target stars in the young open clusters were selected from lists of cluster
  members published by \cite*{pro92} for $\alpha$~Persei, by \cite*{boe87} and
  \cite*{bou93} for Coma Berenices, by \cite*{sch91} for the Hyades,
  by \cite*{sod93a} for the Pleiades, and by \cite*{boe88}, \cite*{snc87} and
  \cite*{sod93b} for Ursa Major.  Within each cluster, preference was given to
  targets with relatively low cirrus confusion, as determined from the IRAS
  Sky Survey background measurements.  In Coma~Berenices, a sparse cluster
  with few confirmed members and relatively low cirrus, we included nearly all
  dwarf stars with spectral types later than A0.  By including all possible
  Ursa Major candidates not reserved by ISO core programs, we sampled a wide
  range of spectral types in this cluster, from early A to early K, although
  most objects are of spectral types F and G.  For the remaining clusters
  where there were larger numbers of candidates available, we gave preference
  to stars with spectral types near solar, mostly late F- and G-type stars.
  When choosing targets in the Hyades, we also took into account indications
  of potential far infrared excesses based on IRAS 12 and 25\,$\micron$ data.
  Binarity was not a cause for exclusion from our sample because we hoped to
  increase understanding of the relationship between circumstellar dust and
  multiplicity.  Finally, we made efforts to include targets with a range of
  rotational velocities to assess the relationship between the presence of a
  disk and stellar rotation.

  Very young stars included CTTs and WTTs in the Chamaeleon~I association
  selected from the published lists of \cite{gau92} and \cite{alc95}, WTTs in
  Taurus from \cite{wic96}, and WTTs in the Upper Scorpius association from
  \cite{wal94}.  In Chamaeleon~I, most sources have measured IRAS fluxes at
  12 and/or 25\,$\micron$ but lack good quality measurements at 60 and/or
  100\,$\micron$.  Other WTTs were selected to span the range of the limited
  age estimates available, and for lower cirrus confusion.

  We also included 19 field stars (hereafter referred to as Young Field Stars)
  which exhibited indications of youth, primarily based on published
  chromospheric activity age indicators.  We intended to limit distances to
  less than 60~pc, but following the publication of Hipparcos data it became
  clear that two observed stars are actually more distant than this cutoff.
  Like the cluster stars, the Young Field Stars span the age range 60 to
  800~Myr with a median of 180 Myr.  They encompass spectral types A through K,
  but are concentrated on spectral types F and G.  Individual stellar ages are
  presented in Table~\ref{tab:yfsa}.

\placetable{tab:yfsa}

\section{Observations and Results \label{sec:obsres}}

  We selected the ISOPHOT C100 detector \citep{lem96} for our target
  observations.  Targets were observed throughout the ISO mission, and our
  observing strategies changed as our understanding of ISOPHOT performance
  improved.  The first group of 63 observations was made between 1996~June and
  1996~December.  The majority employed the 60 and 100\,$\micron$ filters and
  the triangular chopping mode with a 150~arcsec throw.  Eight stars in
  Coma~Berenices (see Table~\ref{tab:sourcesc}) were observed with a
  ``staring chain'' -- an on-target observation followed by a background
  observation 3~arcmin north of the target, then the next on-target
  observation, etc., first in the 60\,$\micron$ filter, then in the
  100\,$\micron$ filter.  The second group of 85 observations was made
  between 1997~June and 1998~March using the 60 and 90\,$\micron$ filters
  and the raster mode in a 3$\times$3 grid with 46~arcsec steps.  A list of
  total integration times and number of targets observed in each cluster is
  presented in Table~\ref{tab:clus}.

  Data reduction made use of the ISOPHOT Interactive Analysis (PIA) package
  version 7.3 \citep{gab97} and was supported by our own routines.  Our final
  results are based on data from Off-Line Processing (OLP) version 7.0.
  Standard reduction parameters were used for linearization of ramps,
  2-threshold deglitching, first-order polynomial voltage ramp fitting, and
  reset interval correction.  Slope deglitching was slightly modified by using
  a sigma range of 2.4 and max/min clipping.  Chopped observations were
  corrected with the default chopper-loss factors given by the ISOPHOT Team
  \citep{gab97}.  Dark-current subtraction employed the orbital-dependent
  model values.  Finally, to limit the influence of the non-linear response of
  the detector to a change in flux, we discarded the first third of each
  raster pointing.

  We calibrated the data with the on-board Faint Calibration Source (FCS)
  measurements.  Since there is significant evidence for detector drifts
  during observation, the best strategy obtains calibration measurements
  immediately before and after a target observation.  For the staring and
  chopped observations, which were all made before the detector drift was well
  understood, FCS measurements were made at the beginning and end of a group
  of individual observations.  Calibration factors were then interpolated for
  each target and background observation.  For raster observations, FCS
  measurements were made immediately before and after each target observation
  and calibration factors interpolated at each raster step.  The application
  of the FCS calibration also removes a first order flat-field.  Residual
  effects exist, but in general their magnitude is less than the other
  observational uncertainties.  As far as possible, these FCS-based
  calibrations were cross-checked by comparing the method of calibration from
  the observed ISO and published IRAS backgrounds and a number of actual IRAS
  source measurements as described by \cite*{sil98}, \cite{sil00} and
  \cite{spa99} and were found to be consistent.  A more detailed discussion of
  the calibration comparisons is given in Appendix~\ref{sec:fluxcal}.  We are
  confident the PIA calibration using the FCS measurements is sufficiently
  reliable to support our conclusions.  Corrections for the point-spread
  function (PSF) were applied using model calculations which give the fraction
  flux falling on the center pixel of the array as 0.66, 0.57 and 0.54 at 60,
  90 and 100\,$\micron$, respectively \citep{lau99}\footnotemark.  A more
  detailed discussion of the standard data reduction procedures is presented
  by \cite{sil00}.

  \footnotetext{ Article available at http://www.iso.vilspa.esa.es/users/expl\_lib/PHT\_list.html}

  For all observing strategies used here, a source was considered detected if
  the flux at the on-target (center) pixel exceeded three times the standard
  deviation of the flux on the eight surrounding pixels.  This provided a
  conservative approach to problematic observations.  In all cases, the
  standard deviation of the flux on the surrounding pixels is higher than any
  other statistical uncertainties associated with the observation.  In five
  cases -- all in the Chamaeleon and Scorpius clusters -- the source was
  extended to one outer pixel.  This pixel was excluded from the calculation
  of the background standard deviation.   At both 60\,$\micron$ and
  100\,$\micron$, typical 3$\sigma$ noise limits for chopped observations were
  100~mJy in 256~s.  For raster observations, corresponding limits were 50~mJy
  at 60\,$\micron$ and 40~mJy at 90\,$\micron$ in 473~s.  In regions of very
  low cirrus, the noise limits were as much as a factor of three lower.
  Conversely, in regions with high cirrus confusion, the noise limits were as
  much as a factor of five higher.  These sensitivities are much worse than
  originally anticipated and, as a result, the detection of excesses was
  somewhat compromised.  A complete list of all observed targets with signals
  and uncertainties is presented in Table~\ref{tab:sourcesc} and
  Table~\ref{tab:sourcesr}.

  A total of 36 targets were detected out of 148 observed.  One-third of the
  detections are in the young Chamaeleon cluster.

\placetable{tab:sourcesc}
\placetable{tab:sourcesr}

\section{IR Excesses \label{sec:excess}}

  In order to determine if there is excess emission at the wavelengths
  observed, a good estimate of the stellar photospheric flux is required.
  For non-PMS cluster members more distant than the Hyades (46~pc), the
  photospheric contributions at 60, 90 and 100\,$\micron$ are negligible
  compared to the measured ISOPHOT sensitivities
  (see Section~\ref{sec:obsres}).  For stars in the Hyades and closer and PMS
  stars, we estimated the photospheric flux from published K-band magnitudes
  \citep*{alc95,gau92,law96,ran96,wal94}.  Several K-band magnitudes were also
  obtained for us by J. Hare (1997, private communication).  Where K-band
  magnitudes were unavailable, we used V-band magnitudes and published V-[12]
  colors \citep{coh87}.  A comparison of the results of these two
  extrapolation methods suggests the uncertainty in the predicted photospheric
  values is less than 10\%, significantly less than the noise in our
  measurements.

  Of our 36 detected sources, 33 show evidence of excess far-infrared
  emission.  Basic characteristics of these excess sources including
  spectral type, V-band magnitude, B-V color, effective temperature,
  rotational velocity, distance, age and multiplicity are listed in
  Table~\ref{tab:data1}.

\placetable{tab:data1}

  About one-third of these sources are in young star forming regions and,
  based on IRAS results, were suspected to have infrared excesses.  For the
  most part, only the 12 and 25\,$\micron$ IRAS measurements are reliable and
  our new measurements at 60 and 100\,$\micron$ complement these.  Our ISO
  observations demonstrate for the first time that there is excess emission at
  60\,$\micron$ and 90 or 100\,$\micron$ for thirteen cluster stars, five Young
  Field Stars and one other field star.  These excess sources are presented in
  Tables \ref{tab:data2a} (CTTs and WTTs) and \ref{tab:data2b} (cluster and
  field stars) together with IRAS measurements and upper limits.

\placetable{tab:data2a}
\placetable{tab:data2b}

  The detections of infrared excess are associated with stars of a range of
  spectral types and luminosities.  For convenience we adopt a parameter that
  is independent of luminosity to describe systems exhibiting excess emission
  from circumstellar dust.  The fractional excess luminosity,
  $f_{\rm{d}} \equiv L_{\rm ex}/L_\star$, where $L_{\rm ex}$ is the luminosity
  of dust and $L_\star$ is the stellar bolometric luminosity, effectively
  provides a measure of the relative dust mass for small particles of radius
  a~$\lesssim 1 \micron$ in systems with roughly the same dominant
  temperature.  Values of $f_{\rm{d}}$ for each of our excess sources were
  calculated following the method of \cite{bac87}, summing the luminosities
  in each wavelength band and including a correction to account for excess
  flux from wavelengths longer than the 90 or 100\,$\micron$ band:
  $f_{\rm{d}} = [ L_{\rm ex(12)} + L_{\rm ex(25)} + L_{\rm ex(60)} + L_{\rm ex(90\,or\,100)} + c ]/L_\star$.

  IRAS 12 and 25\,$\micron$ measurements, where available, were used to
  determine $L_{\rm ex(12)}$ and $L_{\rm ex(25)}$.  However, for the vast
  majority of our targets older than $\sim 20$~Myr, there was either no
  measurable excess emission at 12\,$\micron$ and 25\,$\micron$ or no flux
  measurements available.  Values for $L_{\rm ex(60)}$ and
  $L_{\rm ex(90\,or\,100)}$ were derived from the data in
  Tables~\ref{tab:data2a} and \ref{tab:data2b}.  To account for any
  unmeasured excess emission at wavelengths shorter than 60\,$\micron$ and
  longer than 100\,$\micron$, we used the values of $f_{\rm{d}}$ and the
  25, 60 and 100\,$\micron$ measurements for 14 stars from Table~X in
  \cite{bac93} and calculated the correction factor necessary to reproduce the
  final $f_{\rm{d}}$s from only the 60 and 100\,$\micron$ excesses.
  This correction factor, $c = 1.85$, was then applied where we had only
  60\,$\micron$ and 90 or 100\,$\micron$ data.

\section{Discussion \label{sec:disc}}

\subsection{Cluster Stars}

  We detected at least one infrared excess source in almost every cluster
  observed.  Since the implications of the detections differ by cluster, we
  review each cluster separately.

  Coma Berenices: Three IR-excess sources were detected out of 26 stars
  observed.  While the actual number of detections is small, several
  characteristics of the cluster and the observations combine to make this
  a remarkable result.  Coma Berenices is a very sparse cluster, comprised of
  only 42 optically identified members at the time of our observations, 35 of
  which are of spectral type earlier than K0.  It is one of the oldest open
  clusters we observed, with an age of 500 Myr.  Based on the drop off of
  $f_{\rm{d}}$ with age described below in Section~\ref{sec:atc}, Coma
  Berenices stars should have values of $f_{\rm{d}}$ on average a factor of
  100 less than stars in the $\alpha$~Persei cluster.  The fact that Coma
  Berenices at 88~pc is relatively close increases our sensitivity to excess
  flux there compared to $\alpha$~Persei by a factor of only four.
  Furthermore, this cluster could be observed only early in ISO's lifetime
  when we used the less-sensitive chopping mode with 3$\sigma$ noise limits at
  least four times higher than those of the raster observations.  For the three
  Coma Berenices excess sources to be detectable, they must have values of
  $f_{\rm{d}}$ 10 to 100 times higher than we would expect, indicating they
  are unusual systems.  At least three other stars in the cluster display only
  slightly less than 3$\sigma$ signals on the center pixel, suggesting the
  presence of additional excess systems.  Only very limited information on the
  members of Coma Berenices is available.  Detailed photometric and
  spectroscopic observations and FIR data are nearly non-existent, although
  there are now Two Micron All Sky Survey (2MASS) observations.

  Ursa Major: Our observations provide the first firm evidence for substantial
  excesses around stars in the Ursa Major cluster, which has an age
  $\sim 300$~Myr and lies at a distance $\sim 25$~pc.  This cluster has two
  components: the nucleus and stream \citep*{rom49}.  Searches to date,
  generally involving only the A-type members \citep*{abr98,lec97,skr91},
  produced evidence for only one circumstellar disk -- associated with
  $\beta$~UMa, a nucleus member -- out of 30 separate nucleus and stream
  systems observed.  One of our detected stars, HD~139798, was suggested as a
  25\,$\micron$ excess source by \cite{ste91}, but its status as a possible
  Ursa Major stream member was not noted.  All of our detected excess sources
  are probable members of the Ursa Major stream.  However, all of our
  observations of the nucleus stars were made early in the ISO mission and
  used the chopping mode, with the lower sensitivity as mentioned above.  Two
  nucleus stars exhibited emission at a level just less than our 3$\sigma$
  limit indicating that more sensitive long wavelength observations may reveal
  additional excess systems.

  Pleiades: Here we detected two very bright sources out of fourteen stars
  observed.  While these two excess sources are intriguing they are not likely
  to be typical.  Considering the relatively young age of the Pleiades, just
  over 100~Myr, we might expect a large number of sources with
  $f_{\rm{d}} \sim 10^{-4}$.  However, for a G5 star, the median spectral type
  observed in this cluster, the lowest detectable $f_{\rm{d}}$ was
  $\sim 7\times10^{-4}$.  With decreased sensitivity in many regions of this
  cluster due to high cirrus, it is not surprising we detected only the
  highest-$f_{\rm{d}}$ disks.  We note that, due to ISO's pointing
  constraints, there was very limited time to observe the Pleiades resulting
  in a survey of only a tiny fraction of its members.  The potential for
  detecting additional excess sources in this cluster is high.

  Chamaeleon~I and Upper Scorpius: Perhaps the least surprising of the
  detections are those of the PMS stars in the Chamaeleon~I and Upper Scorpius
  associations.  We were able to detect emission from 14 of 33 observed
  targets.  While the vast majority (13) of the detected stars were known
  previously to be excess sources, most lacked reliable 60 and/or
  100\,$\micron$ fluxes, or were confused with other sources.  Unfortunately,
  the resolution of ISO was insufficient to resolve three of the CTTs we
  observed.  We detected excess in five of the WTTs we observed.  All of the
  non-detected targets are WTTs, and are either highly variable (e.g. VX~Cha)
  or are from the X-ray detected PMS sources \citep{alc95}, many of which were
  determined to be older stars after our program of observations was executed.
  Comments on individual sources are in notes to Table~\ref{tab:data2a}.  Of
  the two stars observed in Upper Scorpius, one resulted in a very strong
  detection, over 100 mJy in both the 60 and 90\,$\micron$ bands.  The
  similarity in appearance to the Chamaeleon sources suggests there are likely
  to be numerous excess sources in this cluster yet to be identified.  In
  fact, \cite{bac98} reported a significant detection of the sum of emission
  from 6 A-type stars in the Sco-Cen association.
  
  $\alpha$~Persei: In contrast to the preceding clusters, the detection rate
  in this cluster was disappointing.  With an estimated age $\sim 50$~Myr,
  $\alpha$~Persei would be expected to contain systems with
  $f_{\rm{d}} \sim 10^{-3}$.  Observations of this cluster are complicated by
  confusion with galactic plane sources as well as a greater distance of
  185~pc.  Nevertheless, an $f_{\rm{d}} \sim {\rm few}\times10^{-4}$ for an F5
  star, the median observed spectral type of this cluster, would be well
  within our detection limits.  However, recent dating of this cluster by
  lithium depletion boundary methods \citep{sta99} yields an age of
  $\sim 125$~Myr, more than double the previous estimate.  This would reduce
  our expected $f_{\rm{d}}$ by a factor of four, putting the brighter cluster
  systems just at our detection limit.  If $\alpha$~Persei is as young as
  first thought, it would be one of the few clusters with ages just older than
  those of the WTTs, such as in Chamaeleon and Taurus, making it extremely
  important for characterizing circumstellar disks at this point in their
  evolution.  More sensitive observations and a resolution of the age
  question are essential for this cluster.

  Taurus: The lack of significant detections in Taurus is also surprising.
  With an average age of about 25~Myr and at 140 pc, Taurus is comparable to
  the Chamaeleon cluster, and thus would be expected to contain a large number
  of detectable excess sources.  But time to observe this cluster was limited
  and thus statistics are poor.  The targets observed were X-ray selected
  WTTs; however, unlike those in Chamaeleon, most Taurus WTTs are confirmed
  PMS stars \citep{bou97}.

  Hyades: This cluster, one of the older clusters at 625~Myr, is relatively
  close at only 46~pc.  This is about twice as far and twice as old as the
  Ursa Major cluster, where our detected $f_{\rm{d}}$s are on the order of
  $10^{-4}$.  For the Hyades we expect $f_{\rm{d}}$ to be on the order of
  $10^{-5}$, right at our detection limit for an F5 star, the median spectral
  type observed in the Hyades.  Thus, typical disks could have escaped
  detection.  Again, observing time on this cluster was limited and we
  observed only a small percentage of cluster members.  More sensitive
  observations are necessary to determine the true incidence of infrared
  excess.

\subsection{Young Field Stars}
 
  For most of these stars, ages were taken from the literature
  (see Table~\ref{tab:yfsa}).  In a few cases, ages were unavailable or
  unconfirmed.  Chromospheric activity is a useful age-indicator for young,
  low-mass stars and can be measured by the brightness of the Ca-II core
  reversal. This reversal is often measured by the so-called S-parameter
  \citep*[and references therein]{dun91,hen96}.  The S-parameters for two
  stars, HD 35850 and HD 209253 were measured with the 0.6-m Coude Auxiliary
  Telescope and the Hamilton Echelle Spectrograph at Lick Observatory by 
  \cite{sil00}.  These data were calibrated by measuring stars in common with
  \cite{dun91} and \cite{hen96}.  Ages (Table~\ref{tab:yfsa}) were
  calculated following the method of \cite{dun91}.
 
  Among the Young Field Stars, we find detectable excesses in seven out of the
  nineteen stars observed.  The values of $f_{\rm{d}}$ for these excess stars
  are comparable to those of some members of the open clusters, but distances
  to these two samples are very different.  On average, the Young Field Stars
  are at 45~pc, and have an age of 200~Myr, making them most analogous to the
  Ursa~Major cluster, where there was also a significant number of detections.
  A greater percentage of Young Field Star observations than the Ursa~Major
  observations used the high-sensitivity raster mode.

  The S-parameter was also measured for the star HD~151044, an Ursa Major
  Stream candidate ($\approx300\times10^{6}$ yrs) \citep{boe88}.  However its
  membership has been questioned on the basis of relatively low chromospheric
  emission and similar but not entirely consistent space motion with the Ursa
  Major Stream \citep{snc87,snm93}.  Our new chromospheric activity age,
  $2.8\times10^{9}$ yrs, is consistent with HD 151044 being too old to be part
  of the Ursa Major Stream.  However, the calculated excess seems extreme if
  the star is so old.

\subsection{The Age--$f_{\rm{d}}$ Correlation \label{sec:atc}}

  As defined above, $f_{\rm{d}}$ is the fractional dust luminosity,
  $L_{\rm ex}/L_{\star}$.  For each cluster and for the Young Field Stars as
  a whole, an average $f_{\rm{d}}$ was calculated as follows: the sum of all
  values for $f_{\rm{d}}$ from detections and non-detections, including
  negative values, was divided by the total number of targets observed.
  Negative values for $f_{\rm{d}}$ result when the observed flux was less than
  the photospheric flux estimate and in all the approximately 40 cases, the
  ``negative excess'' was less than the $3\sigma$ noise level.  The
  Chamaeleon~I sources were separated into two groups because of the wide
  relative age spread of the stars.  The division was set at 10~Myr, resulting
  in groups of nearly equal size.  However, whether Chamaeleon~I is plotted
  as one group or two, the effect on the fitted age--$f_{\rm{d}}$ relation is
  negligible.

\begin{figure}
\plotone{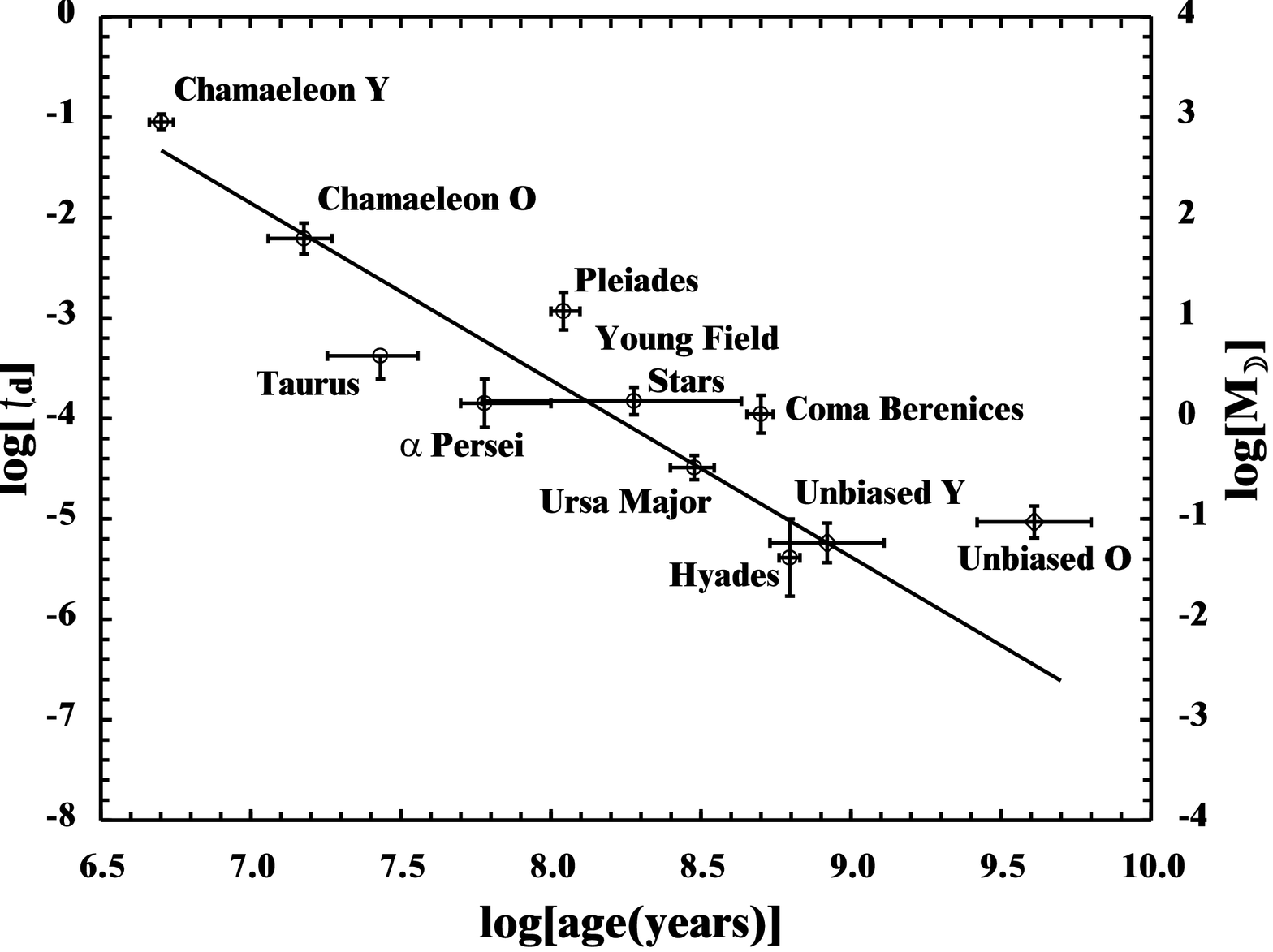}
\caption{\em The behavior of $f_{\rm{d}}$ with stellar age.  Cluster ages are
    taken from the literature (see Table~\ref{tab:clus}).  The plotted line is
    a regression fit to the data with slope~$\approx -1.76y$.  Plotted points
    represent an average f for each cluster that includes both detections and
    non-detections.  Vertical error bars represent the standard deviation of
    the average.  The two points, Unbiased~Y and Unbiased~O are from
    \protect\cite{sil00} and are not included in the calculation of the
    regression.
    \label{fig:agetauC}}
\end{figure}

\begin{figure}
\plotone{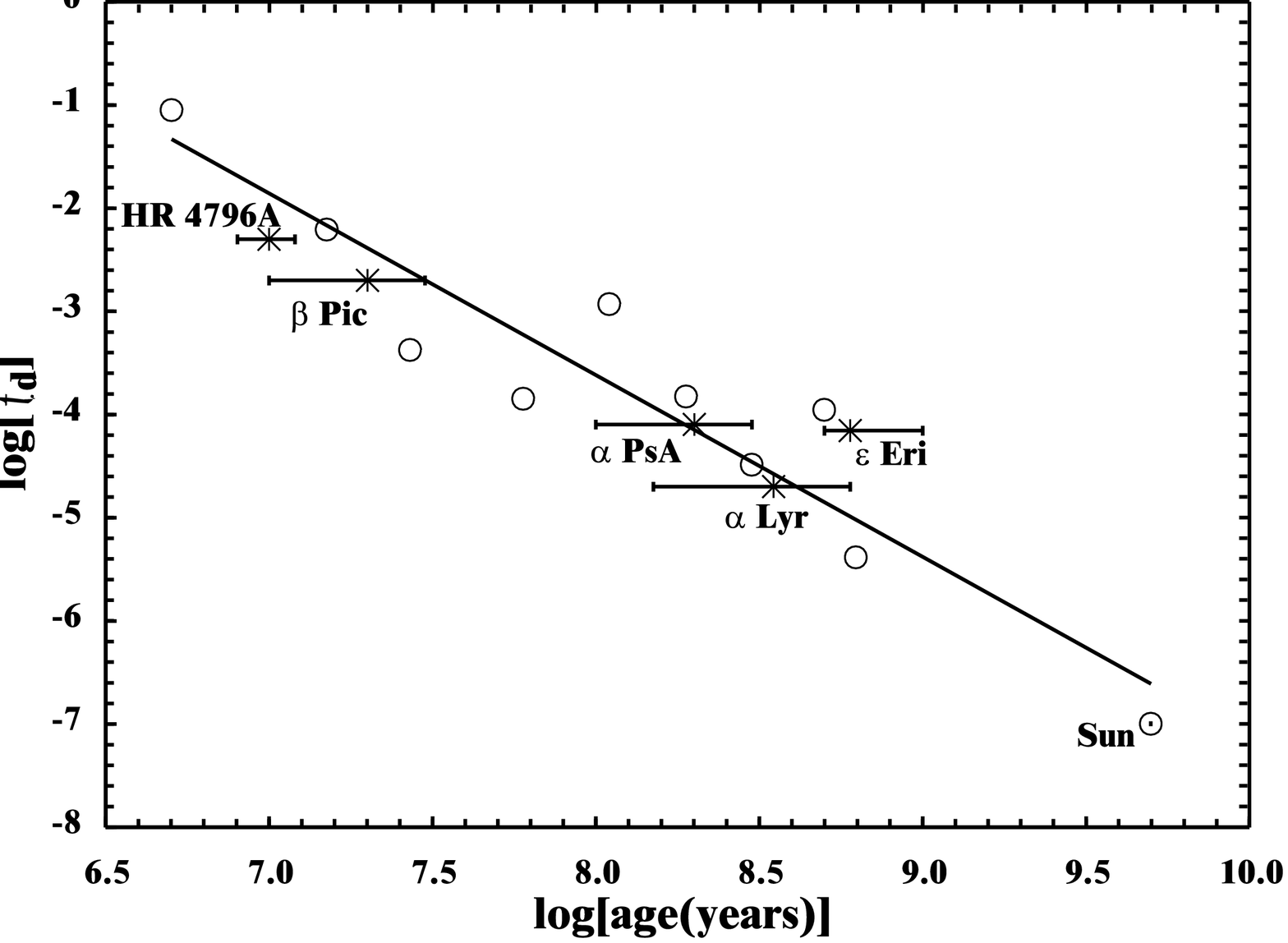}
\caption{\em   Archetypal Vega-phenomenon stars (asterisks) and the Sun are
    over-plotted on the cluster points (open circles) for comparison
    \protect\citep[and references therein]{snd89,jur98,bar99,hol98}.
    The position of the Vega-like stars may be regarded as representing a rough
    upper-envelope of $f_{\rm{d}}$ at a given age.  The value of $f_{\rm{d}}$
    plotted for the Sun is based on emission from Zodiacal dust only
    ($\leq 5$~AU) and does not include the unknown contribution from outside
    the orbit of Neptune which could or could not be significant.
    \label{fig:agetauV}}
\end{figure}

  Figure~\ref{fig:agetauC} shows the averaged $f_{\rm{d}}$s versus the cluster
  ages.  Comparable values for a few nearby Vega-type
  stars are plotted over the main cluster results in Figure~\ref{fig:agetauV}.
  There is clear evidence for a systematic decrease in the fractional
  infrared excess emission with stellar age.  A regression fit to the cluster
  data yields a reasonably good fit with a power-law of the form
  $f_{\rm{d}} \varpropto (\rm age)^{-1.76}$.  This is consistent with
  submillimeter studies of Pleiades, Ursa Major, Taurus and field stars
  by \cite{zuc93}, who found ${\rm Dust Mass} \varpropto (\rm age)^{-2}$.
  In contrast, \citet{hab00} have recently proposed that the majority of
  circumstellar disks disappear after the first 400~Myr based on their
  detection of only a few disks around field stars with ages greater than 
  400~Myr.  While it is difficult to compare the two sets of observations
  directly because of differing target samples and noise levels, we detect
  excess emission from dust around seven stars with ages greater than 400~Myr,
  showing disks do persist beyond 400 Myr though the amount of dust tends to
  decline.

  Since $f_{\rm{d}}$ is a measure of the emission from circumstellar dusk, it
  should be correlated with the mass of dust \cite[see also][]{zuc93}.  
  \cite{sil00} provides an approximate conversion relation, 
  dust~mass~$= f_{\rm{d}} \times 1.4 \times 10^{4}\, {\rm M}_{moon}$,
  where ${\rm M}_{moon}$ is the mass of the moon.  Masses calculated in this
  way lead to the right-hand axis in Figure~\ref{fig:agetauC}.
  
  The power-law index near $-2$ in the age-dust mass relationship can
  possibly be explained by a simple model of collisionally replenished
  secondary dust disks.  If the dust clearing time scale is shorter than the
  age of the system, the amount of secondary dust will be determined by the
  instantaneous collisional rate of large particles.  The typical dust
  clearing time scales are given by \cite{bac93} and \cite{sil00} and are less
  than $10^7$~years.  The collisional rate of large particles dN/dt should go
  as N$^2$ where N is the number density of large planetesimals, comets or
  asteroids in the system.  Thus

  \begin{equation}
  f_d \propto \frac{dN}{dt} \propto -\rm{N}^2.
  \end{equation}

  In addition, it is expected that the loss of large particles will be
  determined by the same collisional type process such that integrating
  equation~1 we find that the number of large particles will decrease with
  time as

  \begin{equation}
  \rm{N} \propto \frac{1}{t}.
  \end{equation}

  Combining equations~1 and 2, we get that the mass of emitting dust is
  given by

  \begin{equation}
  {\rm{M\, (emitting\,\, dust)}} \propto f_d \propto \frac{dN}{dt} \propto \rm{N}^2 \propto \frac{1}{t^2}.
  \end{equation}

  This is similar to the observed power law fall off.  The problem can also
  be reversed.  The fact that we observe the mass of emitting dust dropping
  off as $1/t^2$ suggests that in this model, the number of large particles N,
  in planetesimals, comets and asteroids, falls as 1/t independent of the
  mechanism that produces the fall off.

\placefigure{fig:agetauC}
\placefigure{fig:agetauV}

\subsection{Correlations with other stellar characteristics}

  With only a small number of detections in each cluster, it is difficult to
  draw general conclusions regarding the relationship between circumstellar
  material and stellar characteristics such as multiplicity and rotation.
  We detected dust around single stars, several spectroscopic binaries
  including all three Coma Berenices excess stars and one Young Field Star,
  HD~177996, and one wide binary, HD~125451 (separation $\sim 4000$~AU).
  Therefore, over the age range of 10 to 600 Myr the presence of a stellar
  companion does not necessarily preclude the existence of associated
  circumstellar material. \cite{ost95} and \cite{jen96} show that disks may
  surround members of binary systems in cases where the separation is either
  less than 1~AU, or greater than 100~AU.

  The detected sources exhibit a wide range of rotational velocities.  In the
  young open clusters, half our detected excess sources have {\em v}~sin{\em i}
  less than the median rotational velocity of their parent clusters.  This
  distribution is similar to the {\em v}~sin{\em i} distribution in the
  original sample of targets.  In each cluster about half of the targets had
  {\em v}~sin{\em i} less than the median rotational velocity of the cluster.
  Only in the Pleiades is there a significant imbalance; there the two
  detected sources have the highest {\em v}~sin{\em i} of the stars we
  observed in that cluster, roughly five times the cluster median.  Among the
  Young Field Stars, those detected have a range of rotational velocities
  similar to the young open clusters (including both slow rotators, and stars
  with rotational velocities similar to the two detected Pleiades stars).
  A dearth of comprehensive data for rotational velocities of the PMS stars
  in Scorpius and Chamaeleon~I prevents detailed discussion for these samples.

  Most of our detections in the young open clusters and Young Field Stars
  (i.e. not PMS stars) are stars of spectral type F.  This is partly due to
  selection effects: we observed a large fraction of F-type stars and dust
  around F-type stars would be easier to detect because of the higher
  luminosity than G- or K-type stars.  For F-type stars alone, we detect
  excess IR emission in 12 out of 48 stars observed, or 25\%.  In their
  review, \cite{lag00} state for stars of spectral types A--K the fraction
  with disks is 15\%.  While the fraction for F-type stars specifically is
  not noted, our results suggest the fraction is indeed significantly higher
  than 15\%.  Our results also indicate that future work is likely to confirm
  a similarly higher percentage for G- and K-type stars.

\section{Summary \label{sec:sum}}

  A survey of almost 150 pre-main sequence and young main sequence stars, the
  majority of which are members of young open clusters, has been carried out
  with ISOPHOT.  Excess emission at 60, 90, and/or 100\,$\micron$, presumably
  from circumstellar debris disks, has been detected from 33 of the observed
  stars.  The detections are distributed among the observed clusters/groups.
  Twenty of the stars, most of which are of spectral type F, were not
  previously known to exhibit excess emission, and represent a new set of
  Vega-type systems.  Other ISO programs of observations with similar
  objectives to ours \citep{robb99,hab00} have resulted in much lower
  detection rates.  Much of this can be attributed to greatly different
  observing strategies, including shorter integration times, method of
  measuring background, and choice of targets.  Because of the differences in
  observational techniques, it is difficult to compare the results.

  A major conclusion from our observations is how the excess IR emission
  evolves with time.  The ratio of excess IR luminosity to stellar luminosity,
  $f_{\rm{d}}$, appears to drop off with age according to the power law
  $f_{\rm{d}} \varpropto (\rm age)^{-1.76}$.  A power law
  $\propto (\rm age)^{-2}$ is expected for collisionally replenished secondary
  dust disks.  We do not see evidence for an abrupt cessation of the
  debris disk phenomenon as reported by \cite{hab00}.

  The detection of the new IR excess sources reported here is only the
  beginning.  More work needs to be done to help characterize the spectral and
  spatial extent of systems which have been detected.  We have no
  100\,$\micron$ measurements for any of the young cluster stars which we
  observed in the chopping mode.  Many of the PMS objects we measured at
  100\,$\micron$ are not well characterized spatially.  We expect that the
  DEep Near Infrared Survey of the Southern Sky (DENIS) and 2MASS searches
  will add near-IR data to many of the young cluster sources, but may not be
  able to measure some of the more distant stars.  Observations of our new
  excess sources are also needed in the 10 -- 60\,$\micron$ region.  However,
  the far infrared wavelengths at which dust disks are most readily studied
  are inaccessible from the ground, requiring future high-altitude and space
  missions, such as SOFIA and SIRTF.  Finally, the multiplicity of a number of
  these sources, both young cluster and PMS, needs to be better established.

  Our results indicate that there is great potential for more sensitive
  space-based studies of clusters to identify additional solar-type stars with
  far-infrared excess.  The detection and analysis of many more of these
  intriguing objects will provide a better understanding of the planet
  formation process.

\acknowledgements
  The ISOPHOT data presented in this paper were reduced using PIA, which is
  a joint development by the ESA Astrophysics Division and the ISOPHOT
  consortium.  This work was facilitated by the use of the Vizier service and
  the SIMBAD database developed at CDS.  This work has made use of XCATSCAN,
  an internet service provided by the Infrared Processing and Analysis Center
  (IPAC), which is operated by the Jet Propulsion Lab and the California
  Institute of Technology under contract to NASA.  We would like to thank
  M. Jura and P. Goldreich for assistance with the original ISO program, and
  D. Backman for helpful comments.  This research was supported in part by
  JPL Subcontract No. 1201061 pursuant to JPL NASA Prime Contract Task Order
  No. N1260.

\appendix

\section{Flux Calibration \label{sec:fluxcal}}

\begin{figure}
\figurenum{A1}
\plotone{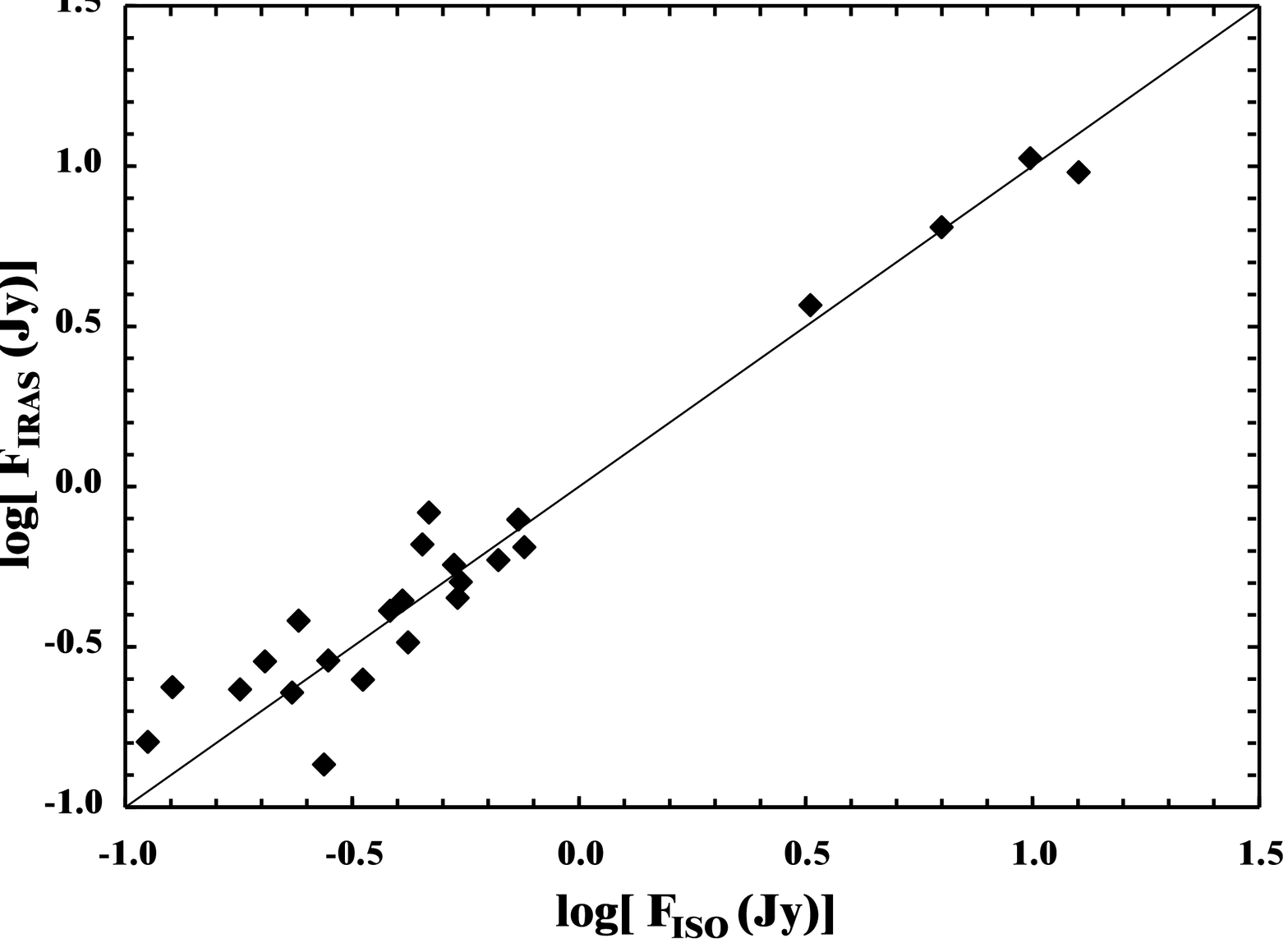}
\caption{\em  Comparison of our calibrated ISO 60\,$\micron$ observations to
              IRAS FSC flux densities.  The line
              ${\rm F}_{\rm ISO} = {\rm F}_{\rm IRAS}$ is plotted for
              reference.
              \label{fig:fluxcal}}
\end{figure}

  As discussed by \cite{spa99} and \cite{sil00}, a number of targets from our
  entire ISO:DEBRIS program had existing IRAS Faint Source Catalogue (FSC)
  measurements at 60\,$\micron$ that were confirmed to be stellar in origin.
  We made use of these data to check the PIA/FCS calibration of our ISOPHOT
  fluxes.  The comparison of 24 sources is plotted in Figure~\ref{fig:fluxcal}
  along with the line ${\rm F}_{ISO} = {\rm F}_{IRAS}$ for reference, and
  shows good agreement between the IRAS FSC and ISO FCS flux measurements.  

\placefigure{fig:fluxcal}

\clearpage

\clearpage

\begin{deluxetable}{cccccccc}
  \tabletypesize{\scriptsize}
  \tablecolumns{8}
  \tablewidth{470pt}
  \tablecaption{Properties of observed clusters. \label{tab:clus}}
  \tablehead{
    \colhead{Cluster}     & \colhead{Age}            & \colhead{ref}         &
    \colhead{Distance}    & \colhead{ref}            & \colhead{Number of}   &
    \colhead{Observation} & \colhead{Integration Time} \\
    \colhead{Name}        &                          &                       &
                          &                          & \colhead{Targets}     &
    \colhead{Mode}        & \\
                          & \colhead{(Myr)}          &                       &
    \colhead{(pc)}        &                          &                       &
                          & \colhead{(s)} \\
  }
  \startdata
    Upper Scorpius &  1 -- 10 & 1 & 145 & 1 &  2 & raster  & 53 / raster pt., 473 total \\
    Chamaeleon I a &  1 -- 20 & 2 & 140 & 3 & 18 & chopped & 128 on-target, 128 off-target \\
    Chamaeleon I b &  3 -- 40 & 4 & 150 & 4 & 11 & raster  & 40 / raster pt., 346 total \\
       & & & & & 2 & raster & 53 / raster pt., 473 total \\
    Taurus         & 10 -- 40 & 5 & 140 & 6 &  8 & raster  & 53 / raster pt., 473 total \\
    $\alpha$ Persei & 50 & 7 & 185 & 8 & 6 & raster & 40 / raster pt., 346 total \\
       & & & & & 13 & raster & 53 / raster pt., 473 total \\
    Pleiades & 120 & 7 & 118 & 8 & 14 & raster & 53 / raster pt., 473 total \\
    Ursa Major & 300 & 9 & 10 -- 40 & 10 & 13 & chopped & 128 on-target, 128 off-target \\
       & & & & & 6 & raster & 53 / raster pt., 473 total \\
    Coma Berenices & 500 & 11 & 88 & 11 & 18 & chopped & 128 on-target, 128 off-target \\
       & & & & & 8 & staring & 128 on-target, 128 off-target \\
    Hyades & 625 & 12 & 46 & 12 & 9 & raster & 53 / raster pt., 473 total \\
    Young Field & 60 -- 630 & 13 & 25 -- 125 & 10 & 6 & chopped & 128 on-target, 128 off-target, 60\,$\micron$ \\
    Stars & & & & &   &       & 64 on-target,  64 off-target, 100\,$\micron$ \\
          & & & & & 13 & raster & 46 / raster pt., 410 total \\
  \enddata
  \tablerefs{ (1) \cite{pre99}, (2) \cite{law96},  (3) \cite{fei93},
              (4) \cite{alc97}, (5) \cite{bou97},  (6) \cite*{ken94},
              (7) \cite*{mey93}, (8) \cite{robi99}, (9) \cite{egg73},
              (10) HIPPARCOS,   (11) \cite*{ode98}, (12) \cite{per98},
              (13) see Table~\ref{tab:yfsa}}
\end{deluxetable}

\clearpage

\begin{deluxetable}{lrclrc}
  \tabletypesize{ \scriptsize }
  \tablecolumns{6}
  \tablewidth{470pt}
  \setlength{\tabcolsep}{0.3in}
  \tablecaption{Young Field Star ages. \label{tab:yfsa}}
  \tablehead{
    \colhead{Object} & \colhead{Age} & \colhead{Ref} &
    \colhead{Object} & \colhead{Age} & \colhead{Ref} \\
                     & \colhead{(Myr)} &               &
                     & \colhead{(Myr)} &               \\
  }
  \startdata
    HD    105 & 500 & 1 & HD 160934 &     &   \\
    HD   1405 &  60 & 2 & HD 171488 &  80 & 6 \\
    HD  16884 & 800 & 3 & HD 175897 & 100 & 1 \\
    HD  35850 & 230 & 4 & HD 177996 & 300 & 1 \\
    HD  36705 &     &   & HD 180445 & 160 & 1 \\
    HD  37484 &  80 & 3 & HD 197890 & 300 & 7 \\
    HD  54579 & 160 & 1 & HD 202917 & 180 & 1 \\
    HD 119022 & 160 & 1 & HD 209253 & 400 & 4 \\
    HD 129333 & 300 & 5 & HD 220140 &  60 & 8 \\
    HD 134319 & 500 & 5 &           &     &   \\
  \enddata 
  \tablerefs{ (1) \cite{hen96}, (2) \cite{gri92}, (3) \cite{fav93}
              (4) This work.    (5) \cite{dun91}, (6) \cite{hen95}
              (7) \cite{jef95}, (8) \cite{man92} }
\end{deluxetable}

\clearpage

\begin{deluxetable}{lcrrrrlcrrrr}
  \tabletypesize{ \scriptsize }
  \tablecolumns{12}
  \tablewidth{470pt}
  \tablecaption{Targets observed in chopped mode with fluxes and
                uncertainties. \label{tab:sourcesc}}
  \tablehead{
    \colhead{Target} & \colhead{Spectral}       & \colhead{60\,$\micron$} &
    \colhead{RMS}    & \colhead{100\,$\micron$} & \colhead{RMS}           &
    \colhead{Target} & \colhead{Spectral}       & \colhead{60\,$\micron$} &
    \colhead{RMS}    & \colhead{100\,$\micron$} & \colhead{RMS}  \\
    \colhead{Name}   & \colhead{Type}           & \colhead{(Jy)}          &
    \colhead{{Jy}}   & \colhead{(Jy)}           & \colhead{(Jy)}          &
    \colhead{Name}   & \colhead{Type}           & \colhead{(Jy)}          &
    \colhead{{Jy}}   & \colhead{(Jy)}           & \colhead{(Jy)} \\
  }
  \startdata
    \sidehead{\em CHAMAELEON I} \\ [-10pt]
    Sz 4         & CTTS &  0.219 & 0.038 &  0.210 & 0.060 &  Glass Ia     & WTTS & 12.970 & 0.236 &  7.216 & 0.445 \\
    SZ Cha       & CTTS &  3.545 & 0.177 &  2.546 & 0.400 &  HM 19        & CTTS & -0.193 & 2.360 & -0.572 & 3.176 \\
    TW Cha       & CTTS &  0.528 & 0.078 &  0.267 & 0.074 &  VX Cha       & WTTS &  0.030 & 0.063 & -0.155 & 0.193 \\
    HM 5         & WTTS &  0.020 & 0.055 &  0.080 & 0.156 &  WX Cha       & CTTS &  0.232 & 0.045 &  0.192 & 0.134 \\
    CED 110      & WTTS &  7.943 & 0.128 & 17.995 & 8.806 &  GK-1         & CTTS &  0.566 & 2.244 &  2.001 & 5.104 \\
    UX Cha       & WTTS &  0.024 & 0.023 &  0.083 & 0.091 &  WY Cha       & CTTS &  0.303 & 0.165 &  0.832 & 0.548 \\
    UZ Cha       & WTTS &  0.268 & 0.081 &  0.229 & 0.093 &  CHX 18       & WTTS &  0.530 & 0.085 &  0.372 & 0.118 \\
    Lk Ha 332-17 & CTTS &  6.715 & 1.460 &  8.338 & 5.474 &  HM 32        & CTTS &  0.462 & 0.029 &  0.353 & 0.077 \\
    Sz 23        & WTTS &  0.575 & 3.033 &  0.183 & 6.940 &  HM Anon      & WTTS &  0.831 & 0.104 &  0.793 & 0.269 \\
    \sidehead{\em URSA MAJOR} \\ [-10pt]
    HD 109011 & K2 V & 0.019 & 0.095 & -0.037 & 0.064 &  HD 124674 & F1 V & 0.050 & 0.069 & -0.101 & 0.100 \\
    HD 109647 & K0   & 0.052 & 0.066 &  0.089 & 0.032 &  HD 125451 & F5 IV & 0.171 & 0.018 &  0.004 & 0.045 \\
    HD 110463 & K3 V & 0.059 & 0.111 & -0.047 & 0.024 &  HD 139798 & F2 V & 0.067 & 0.022 &  0.007 & 0.049 \\
    HD 111456 & F5 V & 0.069 & 0.040 &  0.038 & 0.070 &  HD 141003 & K3 V & 0.502 & 0.326 &  0.208 & 0.212 \\
    HD 113139 & F2 V & 0.025 & 0.044 &  0.016 & 0.037 &  HD 147584 & G0 V & 0.099 & 0.072 &  0.128 & 0.103 \\
    HD 115043 & G2 V & 0.054 & 0.023 &  0.024 & 0.025 &  HD 180777 & A9 V & 0.014 & 0.053 &  0.061 & 0.080 \\
    HD 116656 & A2 V & 0.171 & 0.022 &  0.016 & 0.022 & & & & & & \\
    \sidehead{\em COMA BERENICES} \\ [-10pt]
    HD 	105805 & A3 V &  0.034 & 0.045 &  0.011 & 0.019 &  HD  107793\tablenotemark{a} & F8 V &  0.011 & 0.018 & -0.007 & 0.008 \\
    HD 	106103 & F5 V &  0.005 & 0.041 &  0.003 & 0.008 &  HD  107887\tablenotemark{a} & F5 V & -0.009 & 0.033 &  0.010 & 0.046 \\
    HD 	106691 & F3 V &  0.091 & 0.091 &  0.022 & 0.029 &  SAO  82286\tablenotemark{a} & G1 V &  0.027 & 0.023 &  0.008 & 0.014 \\
    HD 	106946 & F2 V &  0.061 & 0.052 & -0.040 & 0.027 &  HD  107935\tablenotemark{a} & A7 V &  0.085 & 0.063 &  0.050 & 0.052 \\
    HD 	107067 & F9 V &  0.170 & 0.072 &  0.019 & 0.061 &  BD+26 2342 & K0 V &  0.004 & 0.031 &  0.001 & 0.032 \\
    HD 	107132 & F7 V &  0.031 & 0.042 & -0.001 & 0.027 &  HD  108102 & F8 V & 0.150 & 0.036 &  0.031 & 0.033 \\
    HD 	107131 & A7 V &  0.044 & 0.023 &  0.006 & 0.018 &  HD  108154 & F8 V &  0.051 & 0.038 & -0.014 & 0.015 \\
    HD 	107214 & G0 V &  0.105 & 0.055 & -0.030 & 0.026 &  HD  108226\tablenotemark{a} & F6 V &  0.086 & 0.048 & -0.007 & 0.056 \\
    HD 	107399 & G0 V &  0.023 & 0.049 &  0.023 & 0.015 &  Tr     132\tablenotemark{b} & G5 V & -0.048 & 0.035 &  0.063 & 0.034 \\
    HD  107583\tablenotemark{a} & G1 V &  0.061 & 0.078 &  0.000 & 0.018 &  HD  108651 & A0   &  0.177 & 0.055 &  0.022 & 0.041 \\
    HD  107611\tablenotemark{a} & F6 V & -0.009 & 0.031 &  0.013 & 0.039 &  SAO  82335 & G9 V & -0.089 & 0.048 &  0.005 & 0.022 \\
    HD  107685 & F5 V &  0.100 & 0.045 &  0.006 & 0.012 &  HD  108967 & F6 V &  0.051 & 0.038 &  0.042 & 0.019 \\
    HD  107700\tablenotemark{a} & A2 V &  0.067 & 0.014 &  0.025 & 0.029 &  HD  109307 & A4 V &  0.027 & 0.034 &  0.009 & 0.044 \\
     & + G7 III & & & & & & & & & & \\
    \sidehead{\em YOUNG FIELD STARS} \\ [-10pt]
    HD  16884 & K5 V     &  0.137 & 0.070 &  0.002 & 0.046 &  HD  129333 & F8   &  0.018 & 0.026 & -0.020 & 0.043 \\
    HD  36705 & K1 IIIp. &  0.008 & 0.045 &  0.051 & 0.029 &  HD  160934 & K7   & -0.014 & 0.025 & -0.004 & 0.031 \\
    HD 119022 & G2 IV/V  &  0.047 & 0.056 &  0.125 & 0.091 &  HD  175897 & G0 V &  0.043 & 0.054 & -0.041 & 0.031 \\
  \enddata 
  \tablecomments{Spectral type information from SIMBAD.}
  \tablenotetext{a}{Staring Chain observation}
  \tablenotetext{b}{Replace ``Tr'' with ``Cl Melotte 111'' to find in SIMBAD}
\end{deluxetable}

\clearpage

\begin{deluxetable}{lcrrrrlcrrrr}
  \tabletypesize{ \scriptsize }
  \tablecolumns{12}
  \tablewidth{470pt}
  \tablecaption{Targets observed in raster mode with fluxes and uncertainties.
                \label{tab:sourcesr}}
  \tablehead{
    \colhead{Target} & \colhead{Spectral}       & \colhead{60\,$\micron$} &
    \colhead{RMS}    & \colhead{90\,$\micron$}  & \colhead{RMS}           &
    \colhead{Target} & \colhead{Spectral}       & \colhead{60\,$\micron$} &
    \colhead{RMS}    & \colhead{90\,$\micron$}  & \colhead{RMS}  \\
    \colhead{Name}   & \colhead{Type}           & \colhead{(Jy)}          &
    \colhead{{Jy}}   & \colhead{(Jy)}           & \colhead{(Jy)}          &
    \colhead{Name}   & \colhead{Type}           & \colhead{(Jy)}          &
    \colhead{{Jy}}   & \colhead{(Jy)}           & \colhead{(Jy)} \\
  }
  \startdata
    \sidehead{\em UPPER SCORPIUS} \\ [-10pt]
    ScoPMS  60 & WTTS &  0.003 & 0.008 & 0.002 & 0.011 &   ScoPMS 214 & WTTS & 0.116 & 0.008 & 0.129 & 0.034 \\
    \sidehead{\em CHAMAELEON I} \\ [-10pt]
    RX J0850.1-7554 & WTTS & -0.002 & 0.021 &  0.020 & 0.006 &   RX J1035.8-7859 & WTTS &  0.013 & 0.036 &  0.003 & 0.011 \\
    RX J0853.1-8244 & WTTS &  0.026 & 0.019 &  0.000 & 0.007 &   RX J1048.9-7765 & WTTS & -0.029 & 0.030 &  0.007 & 0.011 \\
    RX J0917.2-7744 & WTTS &  0.005 & 0.019 &  0.008 & 0.012 &   RX J1125.8-8456 & WTTS &  0.019 & 0.011 & -0.001 & 0.005 \\
    RX J0928.5-7815 & WTTS &  0.003 & 0.034 &  0.002 & 0.012 &   RX J1203.7-8129 & WTTS & -0.006 & 0.019 & -0.002 & 0.008 \\
    RX J0952.7-7933 & WTTS &  0.005 & 0.022 &  0.003 & 0.009 &   RX J1225.3-7857 & WTTS &  0.004 & 0.009 & -0.003 & 0.009 \\
    RX J1007.7-8504 & WTTS &  0.001 & 0.019 & -0.007 & 0.007 &   RX J1325.7-7955 & WTTS & -0.017 & 0.014 &  0.010 & 0.014 \\
    RX J1009.6-8105 & WTTS &  0.002 & 0.018 & -0.004 & 0.009 & & & & & & \\
    \sidehead{\em TAURUS} \\ [-10pt]
    RX J0408.2+1956 & WTTS &  0.000 & 0.009 &  0.009 & 0.009 &   RX J0420.4+3123 & WTTS &  0.012 & 0.040 &  0.006 & 0.016 \\
    RX J0409.2+2901 & WTTS &  0.011 & 0.032 & -0.005 & 0.021 &   RX J0423.7+1537 & WTTS &  0.029 & 0.026 &  0.006 & 0.018 \\
    RX J0412.8+2442 & WTTS & -0.010 & 0.013 &  0.011 & 0.015 &   RX J0431.3+2150 & WTTS & -0.013 & 0.021 & -0.003 & 0.015 \\
    RX J0415.3+2044 & WTTS &  0.010 & 0.014 & -0.010 & 0.019 &   RX J0457.0+1517 & WTTS & -0.002 & 0.015 &  0.016 & 0.009 \\
    \sidehead{\em $\alpha$ PERSEI\tablenotemark{a}} \\ [-10pt]
    HE  061 & F0      & -0.005 & 0.017 & -0.003 & 0.015 &   HE  660 & F4 V & -0.007 & 0.018 & -0.018 & 0.017 \\
    HE  143 & F8 IV-V &	 0.025 & 0.017 &  0.006 & 0.012 &   HE  715 & F5 V &  0.003 & 0.013 &  0.007 & 0.011 \\
    HE  299 & F6 V    &  0.014 & 0.016 &  0.011 & 0.014 &   HE  733 & F6 V &  0.009 & 0.019 &  0.012 & 0.018 \\
    HE  361 & F4 V    &  0.036 & 0.010 &  0.017 & 0.033 &   HE  750 & F9 V & -0.009 & 0.020 & -0.005 & 0.014 \\
    HE  387 & F2 V    &  0.005 & 0.009 &  0.009 & 0.006 &   AP  166 & G0   &  0.003 & 0.012 & -0.004 & 0.014 \\
    HE  421 & F2 V    &  0.015 & 0.008 &  0.005 & 0.007 &   HE  794 & F6 V & -0.007 & 0.025 &  0.002 & 0.012 \\
    HE  490 & F4 V    & -0.012 & 0.012 & -0.017 & 0.015 &   HE 1101 & G4   & -0.006 & 0.021 & -0.006 & 0.015 \\
    AP   14 & G4      &  0.001 & 0.017 &  0.013 & 0.013 &   HE 1160 & F7   &  0.015 & 0.018 &  0.004 & 0.017 \\
    HE  621 & F4 V    & -0.002 & 0.016 &  0.019 & 0.017 &   HE 1181 & G2   & -0.012 & 0.017 &  0.005 & 0.009 \\
    HE  635 & A9 V    &  0.004 & 0.012 &  0.005 & 0.011 & & & & & & \\
    \sidehead{\em PLEIADES\tablenotemark{b}} \\ [-10pt]
    HII 1132 & F5 V &  0.028 & 0.010 &  0.026 & 0.010 &   HII 2341 & G4 V &  0.006 & 0.027 & -0.003 & 0.009 \\
    HII 1182 & G5 V & -0.011 & 0.011 & -0.006 & 0.005 &   HII 2462 &      &  0.008 & 0.023 &  0.002 & 0.021 \\
    HII 1220 & G8 V &  0.013 & 0.018 & -0.007 & 0.014 &   HII 2506 & F8 V &  0.002 & 0.018 &  0.006 & 0.010 \\
    HII 1593 & G6 V &  0.015 & 0.015 &  0.001 & 0.015 &   HII 2786 & G0   &  0.006 & 0.018 & -0.010 & 0.018 \\
    HII 1924 & G0 V & -0.011 & 0.019 &  0.000 & 0.015 &   HII 3163 & K2   &  0.009 & 0.023 &  0.038 & 0.011 \\
    HII 2284 &      & -0.013 & 0.016 &  0.003 & 0.022 &   HII 3179 & G0 V &  0.019 & 0.031 &  0.012 & 0.005 \\
    HII 2311 &      & -0.013 & 0.021 & -0.006 & 0.024 &   HII 3197 & K3 V & -0.024 & 0.023 &  0.010 & 0.013 \\
    \sidehead{\em URSA MAJOR} \\ [-10pt]
    HD  11131 & G0 V & 0.005 & 0.004 &  0.006 & 0.005 &   HD  72905 & G1 V & 0.077 & 0.014 &  0.053 & 0.012 \\
    HD  13594 & F5 V & 0.000 & 0.004 &  0.009 & 0.009 &   HD 165185 & G5 V & 0.004 & 0.004 &  0.005 & 0.011 \\
    HD  13959 & K4 V & 0.011 & 0.007 & -0.002 & 0.007 &   HD 184960 & F7 V & 0.050 & 0.016 &  0.046 & 0.012 \\
    \sidehead{\em HYADES} \\ [-10pt]
    HD  27429 & F3 V &  0.001 & 0.017 & -0.007 & 0.024 &   HD  28344 & G2 V &  0.016 & 0.026 &  0.007 & 0.018 \\
    HD  27459 & F0 V &  0.050 & 0.013 & -0.004 & 0.012 &   HD  28677 & F4 V & -0.001 & 0.018 &  0.002 & 0.016 \\
    HD  27561 & F5 V &  0.007 & 0.016 &  0.017 & 0.022 &   HD  28992 & G1 V & -0.006 & 0.018 &  0.008 & 0.022 \\
    HD  27848 & F6 V &  0.018 & 0.025 &  0.012 & 0.016 &   HD  29225 & F5 V &  0.009 & 0.022 & -0.009 & 0.016 \\
    HD  28205 & F8 V & -0.012 & 0.015 & -0.003 & 0.005 & & & & & & \\
    \sidehead{\em OTHER} \\ [-10pt]
    HD 151044 & F8 V &  0.088 & 0.0163 & 0.103 & 0.022 & & & & & & \\
    \sidehead{\em YOUNG FIELD STARS} \\ [-10pt]
    HD    105 & G0 V &  0.143 & 0.026 &  0.167 & 0.008 & HD 177996 & K1 V &  0.019 & 0.024 &  0.042 & 0.017 \\
    HD   1405 & G5   & -0.025 & 0.016 & -0.009 & 0.009 & HD 180445 & G8 V &  0.032 & 0.018 &  0.008 & 0.012 \\
    HD  35850 & F7 V &  0.049 & 0.014 &  0.050 & 0.012 & HD 197890 & K0 V &  0.002 & 0.021 &  0.006 & 0.009 \\
    HD  37484 & F3 V &  0.115 & 0.022 &  0.084 & 0.011 & HD 202917 & G5 V &  0.050 & 0.020 &  0.038 & 0.015 \\
    HD  54579 & G0 V &  0.000 & 0.024 & -0.003 & 0.020 & HD 209253 & F6/F7 V &  0.127 & 0.021 &  0.107 & 0.015 \\
    HD 134139 & G5   &  0.048 & 0.019 &  0.012 & 0.011 & HD 220140 & G9 V & -0.001 & 0.011 & -0.006 & 0.013 \\
    HD 171488 & G0 V &  0.032 & 0.024 &  0.017 & 0.018 & & & & & & \\
% HD 181296 &  0.487 & 0.021 &  0.346 & 0.016 \\
  \enddata
  \tablecomments{Spectral type information from SIMBAD.}
  \tablenotetext{a}{names from: HE = \cite*{hec56}, AP = \cite*{sta85}; to
                    find in SIMBAD, replace ``HE'' with ``Cl Melotte 20'' and
                    prepend ``Cl* Melotte 20'' to AP}
  \tablenotetext{b}{names from \cite{her47}; to find in SIMBAD, replace
                    ``HII'' with ``Cl Melotte 22''}
\end{deluxetable}

\clearpage

\begin{deluxetable}{lcrrccrc}
  \tabletypesize{\scriptsize}
  \tablewidth{470pt}
  \tablecaption{Properties of stars with infrared excess.  The list also
                contains the Chamaeleon~I targets observed but not clearly
                detected for comparison.
                \label{tab:data1}}
  \tablecolumns{8}
  \tablehead{
    \colhead{Target}   & \colhead{SpT}       & \colhead{\em m$_{v}$}     &
    \colhead{B-V}      & \colhead{{\em v} sin{\em i}}  &
    \colhead{Distance} & \colhead{Age} & \colhead{Multiplicity} \\
        &    &    &    & \colhead{(km s$^{-1}$)} &
    \colhead{(pc)}     & \colhead{(Myr)} & \\
  }
  \startdata
    \sidehead{\em UPPER SCORPIUS} \\ [-10pt]
     ScoPMS 214 & K0 IV & 11.2 & 1.24 & 90 & 160\phd\phn & 2 & \\ 
    \sidehead{\em CHAMAELEON I: detections} \\ [-10pt]
     Sz 4 &  CTTS & \nodata & \nodata & \nodata & 140\phd\phn & \nodata & \\
     SZ Cha &  CTTS & 11.7 & 1.58 & \nodata & 140\phd\phn & 8 &  Trinary; sep. $\sim 13\arcsec$ and $\sim 5\arcsec$ \\
     TW Cha &  CTTS & 13.3 & 2.02 & \nodata & 140\phd\phn & 20 & \\
     CED 110 &  WTTS & 11.3 & 1.41 & 75 & 140\phd\phn & 5 & \\
     UZ Cha &  WTTS & 14.9 & 1.72 & \nodata & 140\phd\phn & \nodata & \\
     Lk H$\alpha$ 332-17 &  CTTS & 10.7 & 1.20 & 30 & 140\phd\phn & 4 &  Binary; sep. $< 5\arcsec$ \\
     Glass Ia &  WTTS & 12.8 & 1.47 & \nodata & 140\phd\phn & 3 &  Glass Ib; sep. $\sim 2\arcsec$ \\
     WX Cha &  CTTS & 14.8 & 1.69 & \nodata & 140\phd\phn & 5 &  Binary; sep. $< 1\arcsec$ \\
     WY Cha &  CTTS & 14.0 & 1.57 & \nodata & 140\phd\phn & 4 & \\
     XX Cha &  WTTS & 15.3 & 1.53 & \nodata & 140\phd\phn & 40 & \\
     CHX 18N &  WTTS & 12.1 & 1.31 & 25 & 140\phd\phn & 8 & \\
     HM Anon &  WTTS & 11.1 & 1.15 & $<16$ & 140\phd\phn & 10 &  Binary; sep. $< 1\arcsec$ \\
     HM 32 &  CTTS & 13.5 & 1.44 & \nodata & 140\phd\phn & 5 & \\
     RX J0850.1-7554 &  WTTS & 10.6 & 0.74 & 45 & 170\phd? & 16 & \\
     \sidehead{\em CHAMAELEON I: non-detected or contaminated stars} \\ [-10pt]
     HM 5 &  WTTS & \nodata & \nodata & \nodata & 140\phd\phn & \nodata & \\
     UX Cha &  WTTS & \nodata & \nodata & \nodata & 140\phd\phn & \nodata & \\
     Sz 23 &  WTTS & \nodata & \nodata & \nodata & 140\phd\phn & \nodata &  One of at least 3 companions \\
     & & & & & & &  to VW~Cha; sep. $\sim 17\arcsec$ \\
     HM 19 &  CTTS & \nodata & \nodata & \nodata & 140\phd\phn & \nodata & \\
     VX Cha &  WTTS & \nodata & \nodata & \nodata & 140\phd\phn & \nodata & \\
     GK-1 &  CTTS & \nodata & \nodata & \nodata & 140\phd\phn & \nodata & \\
    \\
    \sidehead{\em $\alpha$ PERSEI} \\* [-10pt]
     HE 361 & F4 V & 9.7 & 0.43 & 30 & 184.2 & 65 & \\
    \sidehead{\em PLEIADES} \\ [-10pt]
     HII 1132 & F5 V & 9.4 & 0.45 & 40 & 120\phd\phn & 120 & \\*
     HII 3163 & K2   & 12.7 & 0.96 & 60 & 120\phd\phn & 120 & \\*
    \sidehead{\em URSA MAJOR} \\ [-10pt]
     HD  72905 & G1 V & 5.6 & 0.62 & 10 & \phn14.3 & 300 & \\
     HD 125451 & F5 IV & 5.4 & 0.39 & 40 & \phn26.1 & 300 &  Binary; sep. $\sim 160\arcsec$ \\
     HD 139798 & F2 V & 5.8 & 0.35 & \nodata & \phn35.7 & 300 & \\
     HD 184960 & F7 V & 5.7 & 0.48 & \phn7 & \phn25.6 & 300 & \\
    \sidehead{\em COMA BERENICES} \\ [-10pt]
     HD 107067 & F8 & 8.7 & 0.52 & \phn6 & \phn68.8? & 500 & SB? \\
     HD 108102 & F8 & 8.2 & 0.49 & 35 & 107.1 & 500 & SB2 \\
     HD 108651 & A0 & 6.7 & 0.22 & 18 & \phn79.0 & 500 & SB2 \\
    \sidehead{\em HYADES} \\ [-10pt]
     HD  27459 & F0 V & 5.3 & 0.23 & 68 & \phn47.2 & 625 & \\
    \sidehead{\em YOUNG FIELD STARS} \\* [-10pt]
     HD    105 & G0 V    & 7.5 & 0.60 & 13 & \phn40.2 & 500 & \\*
     HD  35850 & F7 V:   & 6.3 & 0.50 & 40 & \phn26.8 & 230 & \\*
     HD  37484 & F3 V    & 7.2 & 0.37 & 42 & \phn59.5 &  80 & \\*
     HD 134319 & G5      & 8.4 & 0.64 & \nodata & \phn44.3 & 500 & \\*
     HD 177996 & K1 V    & 7.9 & 0.86 & \phn4 & \phn31.8 & 300 & SB\tablenotemark{a}\\*
     HD 202917 & G5 V    & 8.7 & 0.65 & 12 & \phn45.9 & 180 & \\*
     HD 209253 & F6/F7 V & 6.6 & 0.46 & 16 & \phn30.1 & 400 & \\
    \sidehead{\em OTHER} \\* [-10pt]
     HD 151044 & F8 V & 6.5 & 0.54 & \phn5 & \phn29.4 & $> 300$ & \\*
  \enddata
  \tablenotetext{a}{from \cite*{sod98}}

\end{deluxetable}

\clearpage

\begin{deluxetable}{lrrrrrrrc}
  \tabletypesize{ \scriptsize }
  \tablewidth{470pt}
  \tablecolumns{9}
  \setlength{\tabcolsep}{0.125in}
  \tablecaption{PMS Stars with infrared excess in Chamaeleon~I and Upper
                Scorpius.  The list also contains the Chamaeleon~I targets
                observed but not clearly detected for comparison.
                \label{tab:data2a}}
  \tablehead{
    \colhead{Target} & \multicolumn{4}{c}{IRAS Fluxes} &
    \multicolumn{2}{c}{ISO Fluxes\tablenotemark{*}} &
    \colhead{\large $f_{\rm{d}}$} & \colhead{Comment} \\
     & \colhead{12\,$\micron$} & \colhead{25\,$\micron$} &
     \colhead{60\,$\micron$} & \colhead{100\,$\micron$} &
     \colhead{60\,$\micron$} & \colhead{100\,$\micron$} & & \\
     & \colhead{(mJy)} & \colhead{(mJy)} & \colhead{(mJy)} & \colhead{(mJy)} &
     \colhead{(mJy)} & \colhead{(mJy)} & \colhead{$\times10^{-4}$} & \\
  }
  \startdata
    Sz 4 & 88 & 142 & 241 & $< 4300$ & 219 & 210 & 240\phd\phn & \\
    SZ Cha & 260 & 1390 & 3680 & 3410 & 3545 & 2546 & 910\phd\phn & 1 \\
    TW Cha & 238 & 405 & 450 & 110\tablenotemark{a} & 528 & 267 & 480\phd\phn & \\
    CED 110 & 336 & 140 & $< 4930$ & 17130 & 7945 & $< 17995$ & 250\phd\phn & 2 \\
    UZ Cha & 112 & 128 & $< 210$ & $< 4650$ & 268 & 229 & 320\phd\phn & \\
    Lk~H$\alpha$ 332-17 & 2200 & 3370 & 6450\tablenotemark{a} & 8550\tablenotemark{a} & 6715 & $< 8338$ & 640\phd\phn & 3 \\
    Glass Ia & 10400 & 14720 & 9570 & $< 10400$ & 12970 & 7216 & 4120\phd\phn & \\
    WX Cha & 500 & 514 & 250\tablenotemark{a} & $< 15000$ & 232 & $\leq 200$ & 2240\phd\phn & \\
    WY Cha & $< 500$ & 477 & 830\tablenotemark{a} & 20610\tablenotemark{a} & $\leq 300$ & $\leq 830$ & 720\phd\phn & 4 \\
    XX Cha & 395 & 560 & 590 & $< 3360$ & $\leq 530$ & $\leq 360$ & 550\phd\phn & 5 \\
    CHX 18N & \nodata & \nodata & \nodata & \nodata & $\leq 200$ & $\leq 300$ & \nodata & 5 \\
    HM Anon & $< 250$ & $< 250$ & 790 & 4640 & 831 & 793 & 130\phd\phn & 6 \\
    HM 32 & 93 & 212 & 660 & 1080\tablenotemark{a} & 462 & 353 & 1010\phd\phn & 7 \\
    RX J0850.1-7554 & $< 150$ & $< 200$ & $< 100$ & $< 200$ & $< 62$ & 20\tablenotemark{b} & 6.9 & 8 \\
    ScoPMS 214 & $< 400$ & $< 450$ & $< 600$ & $< 3000$ & 116 & 129\tablenotemark{b} & 27.1 & 9 \\
  \cutinhead{Non-detected or contaminated PMS stars}
    HM 5 & 270 & 129 & $< 190$ & $< 5390$ & $< 165$ & $< 470$ & \nodata & 10 \\
    UX Cha & 114 & $< 150$ & $< 1020$ & $< 82000$ & $< 70$ & $< 272$ & 90\phd\phn & \\
    Sz 23 & $< 1000$ & $< 1250$ & \nodata & \nodata & $\leq 575$ & $\leq 183$ &\nodata & 11 \\
    HM 19 & \nodata & \nodata & \nodata & \nodata & $\leq 1000$ & $\leq 1550$ & \nodata & 12 \\
    VX Cha & $< 100$ & $< 100$ & $< 200$ & $< 8000$ & $< 190$ & $< 600$ & \nodata & 13 \\
    GK-1 & \nodata & \nodata & \nodata & \nodata & $\leq 5500$ & $\leq 22300$ & \nodata & 14 \\
  \enddata
  \tablenotetext{*}{Photospheric estimates have not been subtracted from the
                    fluxes presented.  Upper limits given are 3$\sigma$.}
  \tablenotetext{a}{From co-added IRAS data and DECON \cite{wea92}}
  \tablenotetext{b}{Raster measurement at 90\,$\micron$}
\end{deluxetable}

\clearpage

\scriptsize

    Notes --

\noindent
    1. \cite{hen93} measured a 1.3mm flux.  We included for the purpose of
       checking calibration of ISO fluxes. \\
    2. Indistinguishable from the nearby reflection nebula in the IRAS beam.
       At 60\,$\micron$, our observations indicate a point source.  At
       100\,$\micron$ we are only able to place an upper limit to the flux. \\
    3. \cite{hen93} measured a 1.3mm flux.  At 60\,$\micron$, our observations
       indicate a point source.  At 100\,$\micron$ the data appear to be
       contaminated by flux from the nearby B-type star HD 97048 and we give
       only an upper limit. \\
    4. Another TTS, CHX 15a falls on one of the source-frame pixels.  IRAS
       detected flux from both objects, and we give an upper limit to the
       contribution from WY Cha. At 60\,$\micron$, the total fluxes agree with
       the IRAS value; at 100\,$\micron$, cirrus confusion is very high. \\
    5. Originally identified as the source of the far-IR emission detected
       by IRAS at this position.  Subsequent X-ray studies questioned this
       association, and \cite{wal92} considered nearby CHX~18N (also WTTS,
       separation $\sim 24\arcsec$) to be the IR source.  In our ISO
       observations, the two stars fall on different pixels in the source
       frame.  Both pixels show detected emission, and it seems that each
       star contributes part of the far-IR emission. \\
    6. Has surrounding nebulosity.  Appears to be a point source at
       60\,$\micron$, but at 100\,$\micron$ appears either extended or
       confused with the nebulosity. \\
    7. \cite{hen93} measured a 1.3mm flux.  Our measurements at 60 and
       100\,$\micron$ are significantly lower than IRAS co-added data.
       Evidence for either extension or nebulosity confusion at
       100\,$\micron$. \\
    8. One of the X-ray selected WTTS \citep{alc95}, and it was one of the few
       confirmed as PMS by detection of lithium \citep{cov97}. \\
    9. An X-ray selected WTTS \citep{wal94}.  It appears quite extended
       in our data at both wavelengths. \\
    10. \cite{har93} notes that this source has essentially disappeared since
        its first detection. \\
    11. IRAS lists no data for either VW Cha or Sz 23 due to the proximity of
        the Infrared Nebula.  Both stars fall on the same ISOPHOT pixel and we
        provide an upper limit to their combined fluxes. \\
    12. Contaminated with flux from nearby PSC 11072-7727, the Infrared
        Nebula. \\
    13. This source appears to be dramatically variable.  While \cite*{pru92}
        give near-IR magnitudes, it went undetected by the recent DENIS
        observations \citep{cam98}. \\
    14. Contaminated with flux from nearby HD 97300. \\

\normalsize

\clearpage

\begin{deluxetable}{lrrrrrrrc}
  \tabletypesize{ \scriptsize }
  \tablewidth{470pt}
  \tablecolumns{9}
  \setlength{\tabcolsep}{0.175in}
  \tablecaption{Open Clusters and Young Field Stars: IRAS PSC and FSC fluxes
                and new photometry of stars with infrared excess.
                \label{tab:data2b}}
  \tablehead{
    \colhead{Target} & \multicolumn{4}{c}{IRAS Fluxes} &
    \multicolumn{2}{c}{ISO Fluxes\tablenotemark{*}} &
    \colhead{\large $f_{\rm{d}}$} & \colhead{Comment} \\
     & \colhead{12\,$\micron$} & \colhead{25\,$\micron$} &
     \colhead{60\,$\micron$} & \colhead{100\,$\micron$} &
     \colhead{60\,$\micron$} & \colhead{90\,$\micron$} & & \\
     & \colhead{(mJy)} & \colhead{(mJy)} & \colhead{(mJy)} & \colhead{(mJy)} &
     \colhead{(mJy)} & \colhead{(mJy)} & \colhead{$\times10^{-4}$} & \\
  }
  \startdata
    \sidehead{\em $\alpha$ PERSEI} \\* [-10pt]
     HE 361 & $< 250$ & $< 250$ & $< 400$ & $< 1800$ & 36 & $< 100$ & 6.2\phn & \\
    \sidehead{\em PLEIADES} \\ [-10pt]
     HII 1132 & 224 & $< 240$ & $< 280$ & $< 2700$ & 28 & 26 & 66.0\phn & 1 \\ 
     HII 3163 & $< 100$ & $< 300$ & $< 300$ & $< 2000$ & $< 70$ & 38 & 93.4\phn & \\
    \sidehead{\em URSA MAJOR} \\ [-10pt]
     HD 72905 & 885 & 208 & $< 180$ & $< 600$ & 77 & 53 & 2.8\phn & \\
     HD 125451 & 717 & 203 & $< 160$ & $< 440$ & 170 & $< 200$\tablenotemark{a} & 0.77 & \\
     HD 139798 & 497 & 127 & $< 240$ & $< 800$ & 62 & $< 150$\tablenotemark{a} & 0.33 & \\
     HD 184960 & 605 & 166 & $< 170$ & $< 1100$ & 50 & 46 & 1.0\phn & \\
    \sidehead{\em COMA BERENICES} \\ [-10pt]
     HD 107067 & $< 110$ & $< 270$ & $< 280$ & $< 700$ & 170 & $< 200$\tablenotemark{a} & 11.5\phn & 2 \\
     HD 108102 & $< 150$ & $< 210$ & $< 240$ & $< 600$ & 150 & $< 200$\tablenotemark{a} & 5.9\phn & \\
     HD 108651 & 154 & $< 200$ & $< 200$ & $< 340$ & 177 & $< 200$\tablenotemark{a} & 1.3\phn & \\
    \sidehead{\em HYADES} \\ [-10pt]
     HD  27459 & 530 & $< 160$ & $< 350$ & $< 1000$ & 50 & $< 36$ & 0.95 & \\
    \sidehead{\em YOUNG FIELD STARS} \\ [-10pt]
     HD    105 & \nodata & \nodata & \nodata & \nodata & 143 & 167 & 2.9\phn & \\
     HD  35850 & 440 & 80 & $< 180$ & $< 1930$ & 49 & 50 & 0.26 & \\
     HD  37484 & 120 & 120 & 130 & $< 410$ & 115 & 84 & 1.6\phn & \\
     HD 134319 & \nodata & \nodata & \nodata & \nodata & 48 & $< 33$ & 1.6\phn & \\
     HD 177996 & 220 & $< 170$ & $< 170$ & $< 830$ & $< 72$ & 42 & 0.47 & \\
     HD 202917 & \nodata & \nodata & \nodata & \nodata & 50 & 38 & 2.5\phn & \\
     HD 209253 & 290 & $< 150$ & 140 & $< 680$ & 127 & 107 & 1.0\phn & \\
    \sidehead{\em OTHER} \\ [-10pt]
     HD 151044 & 321 & 73 & 116 & $< 640$ & 88 & 100 & 1.5\phn & 3 \\
  \enddata
  \tablenotetext{*}{Photospheric estimates have not been subtracted from the
                    fluxes presented.  Upper limits given are 3$\sigma$.}
  \tablenotetext{a}{Chopped upper limit at 100\,$\micron$}
  \tablecomments{
    \\
    1. IRAS FSC has a 12\,$\micron$ measurement that is much too high to be
       photosphere, though there no indication that this has ever been
       identified as an IR excess source.  Our own observations with the
       UCLA double-beam infrared camera at the Lick 3m Shane telescope
       show normal magnitudes at J- and K-bands for this star's
       spectral type. \\
    2. The SB nature of this source is unconfirmed.  \cite{ode98} note that
       the Hipparcos distance for this object is inconsistent with its
       photometric distance and placing the star at the cluster distance
       gives values for its space motion that is more consistent with cluster
       mean values. \\
    3. Once considered an Ursa Major candidate.  Please see text.
  }
\end{deluxetable}

\end{document}